\newif\ifnew\newtrue
\ifnew
     \documentclass[12pt]{article}
  \else
     \documentstyle[12pt]{article}
     \newcommand{\imath}{\i}
     \newcommand{\jmath}{\j}
\fi
\csname @addtoreset\endcsname{equation}{section}
\newcommand{\eqn}[1]{(\ref{#1})}
\newcommand{\ft}[2]{{\textstyle\frac{#1}{#2}}}
\newsavebox{\uuunit}
\sbox{\uuunit}
    {\setlength{\unitlength}{0.825em}
     \begin{picture}(0.6,0.7)
        \thinlines
        \put(0,0){\line(1,0){0.5}}
        \put(0.15,0){\line(0,1){0.7}}
        \put(0.35,0){\line(0,1){0.8}}
       \multiput(0.3,0.8)(-0.04,-0.02){12}{\rule{0.5pt}{0.5pt}}
     \end {picture}}

\newcommand{\Poin}{Poincar\'e}
\newcommand{\beq}{\begin{equation}}
\newcommand{\eeq}{\end{equation}}
\newcommand  {\Rbar} {{\mbox{\rm$\mbox{I}\!\mbox{R}$}}}
\newcommand{\NP}[1]{Nucl.\ Phys.\ {\bf #1}}
\newcommand{\PL}[1]{Phys.\ Lett.\ {\bf #1}}

\newcommand{\be}{\begin{equation}}
\newcommand{\ee}{\end{equation}}
\newcommand{\bea}{\begin{eqnarray}}
\newcommand{\eea}{\end{eqnarray}}
\newcommand{\ba}{\begin{array}}
\newcommand{\ea}{\end{array}}

\renewcommand{\a}{\alpha}
\renewcommand{\b}{\beta}

\newcommand{\vp}{\varphi}
\newcommand{\si}{\sigma}
\def\bbox{{\,\lower0.9pt\vbox{\hrule \hbox{\vrule height 0.2 cm
\hskip 0.2 cm \vrule height 0.2 cm}\hrule}\,}}
\newcommand{\dsl}{\pa \kern-0.5em /}

\newcommand{\pa}{\partial}
\renewcommand{\t}{\theta}
\newcommand{\T}{\Theta}

\newcommand{\nn}{\nonumber\\}

\font\mybb=msbm10 at 12pt
\def\bb#1{\hbox{\mybb#1}}
\def\ie {{\it i.e.}}

\def\bR {\bb{R}}
\def\bE {\bb{E}}

\def\narx {{\vec{\nabla} X}}
\def\piv {\mbox{$\vec{\Pi}$}}

\def\sac{\, , \qquad}

\def\gf{G^{(5)}}
\def\fs{field strength }
\def\wv{worldvolume }
\def\bg{background }
\def\su{supergravity }

\begin{document}
\begin{titlepage}
\begin{flushright}
UB-ECM-PF-99/01\\
KUL-TF-99/5\\
UTTG-01-99\\
hep-th/9901060
\end{flushright}
\vspace{.5cm}
\begin{center}
\baselineskip=16pt
{\Large\bf BPS solutions of a D5-brane worldvolume in a D3-brane background
from superalgebras}
\vskip 0.3cm
{\large {\sl }}
\vskip 10.mm
{\bf  Ben Craps$^{+,1}$,  ~Joaquim Gomis$^{*,\dagger,2}$ }
\\[2mm]
{\bf David Mateos$^{*,3}$
 ~and{}~Antoine Van Proeyen$^{+,4}$  } \\
\vfill
{\small
$^*$
  Departament ECM, Facultat de F\'{\i}sica\\
  Universitat de Barcelona and Institut de F\'{\i}sica d'Altes Energies,\\
  Diagonal 647, E-08028 Barcelona, Spain
\\ \vspace{6pt}
 $^+$
Instituut voor theoretische fysica, \\
Katholieke Universiteit Leuven, B-3001 Leuven, Belgium
\\ \vspace{6pt}
$^\dagger$
Theory  Group, Physics Department\\
University of Texas at Austin\\
Austin, TX 78712 USA
}
\end{center}
\vfill
\par
\begin{center}
{\bf ABSTRACT}
\end{center}
\begin{quote}
The BPS method is used to find BPS solutions of the worldvolume theory of
a
D5-brane in the near horizon geometry of a D3-brane. The BPS
bound is interpreted in terms of the `maximally extended' D5 worldvolume
supersymmetry algebra in
the corresponding curved background, which is $OSp(1|16)$. This algebra is
an extension
of the worldvolume superalgebra
$OSp(4^*|4)$. The analysis is generalized to the non-near horizon case.
\vfill
 \hrule width 5.cm
\vskip 2.mm
{\small
\noindent $^1$ Aspirant FWO, Belgium, E-mail: Ben.Craps@fys.kuleuven.ac.be \\
\noindent $^2$ E-mail: gomis@ecm.ub.es, gomis@zippy.ph.utexas.edu \\
\noindent $^3$ E-mail: mateos@ecm.ub.es \\
\noindent $^4$ Onderzoeksdirecteur FWO, Belgium, E-mail: Antoine.VanProeyen%
               @fys.kuleuven.ac.be  }
\end{quote}
\end{titlepage}
\section{Introduction}
Worldvolume techniques have been a powerful tool in analyzing
configurations of intersecting D-branes \cite {cm, g}
(and M-branes \cite {hlw}). In
\cite{cm, g} a fundamental string ending on a D-brane was analyzed
from the point of view of the D-brane worldvolume theory. That
theory was shown to admit solutions allowing an interpretation as
fundamental strings ending on the brane. This result can be
understood in terms of `central'
charges\footnote{\label{fn:central} We denote by `central' charges
the bosonic charges that appear in the right hand side of the
anticommutator of the fermionic generators apart from the
worldvolume coordinate transformations and $R$-symmetry
generators. The terminology originates from charges allowed in 4
dimensional extended supersymmetry by \cite{HLS}, where these
charges are really central. The anticommutator of the
supersymmetries can contain analogous charges that do not commute
with the Lorentz algebra and/or the $R$-symmetry
\cite{centralhighd,JWvHAVP}.} appearing in the maximally extended
worldvolume supersymmetry algebra \cite{BGT}. In \cite{jqp} these
charges were interpreted along the lines of \cite{bog}. The
Hamiltonian density was written as a sum of squares, such that
putting one of the squares equal to zero yields a first order
differential equation implying the equations of motion.
\par
This worldvolume analysis was performed for branes in a flat
background. The aim of the present paper is to extend the method to the
case of
a brane in the background produced by another brane. We have chosen the
example
of a D5-brane in the background of $N$ D3-branes, since this configuration is
physically particularly interesting. In the near horizon limit of the
D3-brane geometry it is
related to the baryon vertex in ${\cal{N}}=4, D=4$ SYM \cite{witten, Imamura}. In
the
non-near horizon case it is relevant for the Hanany-Witten effect
\cite{hanany,
Callan}.
\par
Two main tools are used in the analysis. First, the Hamiltonian formalism
is
adopted to rederive the BPS equations of \cite{Imamura} and \cite{Callan}
from a bosonic point of view.
Second, this analysis is shown to fit in the description based on
worldvolume
superalgebras.
\par
Section~\ref{review} contains a review of
the worldvolume analysis of a D5-brane in a
flat background.
In Section~\ref{ham.anal.} we write the Hamiltonian density as a sum of squares along
the lines of \cite{jqp}, derive a BPS bound and interpret its
physical meaning.
Section~\ref{algebras} deals with the worldvolume superalgebra
approach, showing how the charges fit in extended worldvolume
algebras.
The conclusions are presented in Section~\ref{ss:conclusions}.
A first appendix contains some notations. A second one gives the
group theoretical structure of the worldvolume superalgebras and
illustrates relations between such superalgebras.
\section{BPS Method for D-branes in a Flat background}

\label{review}
In this section we will review the main points of
the BPS method for D-branes in
a flat background \cite{BGT, jqp}. The relation between the BPS bounds and
the
charges occurring in the worldvolume supersymmetry algebra will be
stressed.
We will work with the particular case of a D5-brane,
since in the next section we will be interested in extending
this analysis to a D5-brane in a non-trivial background.
\par
Let us thus consider a D5-brane in a flat background. The action is
given by
\be
S = - T_5 \int d^{5+1} \sigma \, \sqrt{-\det (g+F)} \ ,
\ee
where $T_5=1/g_s\,{(2\, \pi)}^5\,{\a'}^3$ is the
D5-brane tension (from now onwards, we set $T_5\equiv 1$),
$g$ is the induced metric and $F=dA$ is the field strength of the Born-Infeld
(BI) gauge vector $A$.
Let us fix the static gauge
\be
X^0 = \si^0 \sac X^4=\si^1 , \ldots , X^8=\si^5 \
\ee
corresponding to a brane extending (asymptotically)
along directions 45678.
\par
Now let us look
for `spike-like' static worldvolume solutions describing a
fundamental string attached to the D5.  We consider configurations
with only one scalar excited and with purely electric BI field:
\be
X^1 = X^2 = X^3 = 0 \sac X^9=X(\si) \sac A_t = A_t(\si) \sac
 \vec{A} \equiv (A_4,\ldots ,A_8)=\vec{A}(\si_0) \ .
\label{truncation}
\ee
{}From the 10D space-time point of view one might represent the
configuration we are looking for, by the following array:
\be
\ba{ccccccccccl}
 &&&&&&&&&&\quad \mbox{flat background}      \nn
D5: &\_&\_&\_&4&5&6&7&8&\_&\quad \mbox{worldvolume}      \nn
F1: &\_&\_&\_&\_&\_&\_&\_&\_&9 &\quad \mbox{BPS solution}
\ea
\ee
Since we are interested in bounds on the energy for these configurations,
it is useful to pass
to the Hamiltonian formalism. Let us
denote the canonically conjugate
momenta associated to $X$ and $\vec{A}$ by
$P$ and $\piv$ respectively.
The canonical phase-space Lagrangian density is defined by
\be
{\cal{L}} = \dot{\vec{A}} \cdot \vec{\Pi} + \dot{X} \, P - L =
\vec{E} \cdot \vec{\Pi} + \dot{X} \, P - L - A_t \,
\vec{\nabla} \cdot \vec{\Pi}  \ ,
\ee
where $L$ is the original Lagrangian density, $\vec E=\dot {\vec
A}-\vec\nabla A_t$,
and in the last step we have
integrated $\vec{\nabla}A_t \cdot \vec{\Pi}$ by
parts. In Hamiltonian form $\cal{L}$ is given by
\be
{\cal{L}} =  \sqrt{\left( 1+ (\narx)^2 \right)
\left(1+ P^2 \right)+
 \piv^2 + \left( \piv \cdot \narx \right)^2}
\,\,-\,   A_t \, \vec{\nabla} \cdot \Pi   \ .
\ee
The first term is the desired expression for the energy
density
\be
{\cal{H}}= \sqrt{\left( 1+  (\narx)^2 \right)
\left(1+ P^2 \right)+
 \piv^2 +  \left( \piv \cdot \narx \right)^2}\ ,
\label{hamiltonian}
\ee
whereas the second term yields the Gauss law
(constraint)
\beq
\vec{\nabla} \cdot \vec{\Pi}=0~.
\eeq
For static configurations, $P=0$,
the Hamiltonian reduces to \cite{jqp}
\be
{\cal{H}} = \sqrt{\left( 1 \pm  \narx \cdot \piv \right)^2
+ \left(\narx \mp \piv \right)^2}  \ .
\ee
In order to get a  bound for the energy from the
previous expression we should consider the `vacuum' or
ground state solution
of our flat D5-brane and its energy density.
The ground state solution corresponds to $X(\sigma)$
not being excited and to vanishing BI vector.
This configuration is a solution of the equations of motion and
its energy density is \mbox{${\cal{E}}_{gs}=1$}. Hence, the energy
density in \eqn{hamiltonian} can be split as
${\cal{H}}={\cal{E}}_{gs}+{\cal{E}}_{ren}$,
where ${\cal{E}}_{ren}$ is the renormalized
energy density of the brane, \ie\ the energy density
of the brane relative to its ground state.
After this splitting, we obtain a bound for the energy
density ${\cal{E}}_{ren}$ given by
\be
{\cal{E}}_{ren}\geq  \mid \narx \cdot \piv \mid \ ,
\ee
with equality when
\be
\narx = \pm \piv  \ .
\label{bound}
\ee
The bound on the density implies the following bound on the
renormalized energy:
\bea
E&=&\int_\Sigma \, {\cal{E}}_{ren} \, \geq \, \mid Z_{el} \mid \ , \nonumber\\
Z_{el}&\equiv& \int_\Sigma \, \piv \cdot \narx \ ,
\label{zel}
\eea
where $\Sigma$ is the worldspace of the D5-brane.
\par
Let us now interpret the physical meaning of $Z_{el}$.
Because of the Gauss law constraint, solutions of \eqn{bound}
correspond to solutions of $\nabla^2 X=0$. Isolated singularities of
$X$ are the BIons found in \cite{cm, g}:
\be
X=\frac{q}{V_{(4)} r^3}\ ;\qquad r^2= \left(\si^1\right) ^2 +\ldots +
\left( \si^5\right) ^2\ ;
\qquad V_{(4)}=\frac{8\pi^2}{3}\ ,
\label{Bion}
\ee
corresponding to a charge $q$ at the origin. This charge at the origin
is the source of the BI vector.\footnote{\label{fn:Gauss}The initial generic configuration
had no
source for
the BI vector. The case of a D5 in the background of a D3 studied in the
next section
will be different since for a  generic configuration there is a source for
the
BI vector due to the Wess-Zumino term.}
The Gauss law is only valid away from the
singularity $r=0$, and thus
\be
Z_{el}=q \lim_{r\rightarrow 0} X(r) \ ,
\ee
where $X(r)$  represents the `height' of the spike at a distance
$r$ from the origin.
The charge $q$ should be quantized in the
quantum theory (see {\it e.g.} \cite{cm}).
Setting it equal to its minimum value and
restoring the factor $T_5$ we finally find
\be
Z_{el}= T_f \lim_{r\rightarrow 0} X(r) \ ,
\ee
where $T_f=1/2\,\pi\,\a'$ is the tension of a fundamental string.
This makes clear the interpretation of $Z_{el}$ as the energy
of a fundamental string attached to the D5-brane.
\par
Now let us show how $Z_{el}$ appears as a central charge in the
worldvolume supersymmetry algebra of the D5-brane. In order to construct
this
algebra it is
useful to study first which bosonic background isometries preserve the vacuum
solution of the D5.
The worldvolume part of these symmetries
is $ISO(5,1)$ and the R-symmetry part is given by $SO(4)=SU(2)\times
SU(2)$.
The supersymmetry generators transform in the spinor representation of
this
symmetry algebra, so that they can be denoted as
$Q_\alpha^i$ and $\tilde{Q}_\alpha^{\bar\imath}$,
where $\alpha =1\dots4$ and $i,\bar\imath=1,2$. The `maximally extended'
superalgebra\footnote{In this
algebra one allows the most general anticommutator of the supersymmetries.
This
corresponds
to extending the original symmetry algebra of the D5-brane, which we
call the `worldvolume algebra', to a larger one encoding all BPS
intersections.}
is given by \cite{BPvdS}:
\bea
\{ Q_\a^i , {Q}_\b^j \} &=& \epsilon^{ij}\gamma^\mu_{\a\b}(P_\mu+Z_\mu)
+\gamma^{\mu\nu\rho}_{\a\b}Z^{+(ij)}_{\mu\nu\rho}~; \nn
\{ \tilde{Q}_\a^{\bar{\imath}} , \tilde{Q}_\b^{\bar{\jmath}} \} &=&
\epsilon^{\bar{\imath}\bar{\jmath}}\gamma^\mu_{\a\b}(P_\mu-Z_\mu)
+\gamma^{\mu\nu\rho}_{\a\b}
\tilde{Z}^{-(\bar{\imath}\bar{\jmath})}_{\mu\nu\rho}~; \nn
\{ Q_\a^i , \tilde{Q}_\b^{\bar{\jmath}} \} &=&
{\cal C}_{\a\b}Y^{i\bar{\jmath}}
+\gamma^{\mu\nu}_{\a\b}Y^{i\bar{\jmath}}_{\mu\nu}  \ .
\eea
This algebra is the semi-direct
sum of a contraction of $OSp(1|16)$ with the Lorentz and R-symmetry
algebras; see Appendix~\ref{app:superalgebras} for a related discussion.
\par
The charge $Y^{i\bar{\jmath}}$ is a vector of the R-symmetry group
$SO(4)$. Interpreting this group as the rotation
group of the four-dimensional space transverse to the D5 \cite{BGT}, we see that
$Y^{i\bar{\jmath}}$ selects a direction in this transverse space,
as a string ending on the D5 does.
Hence the charge $Y^{i\bar{\jmath}}$ is associated to
fundamental strings ending on the D5-brane. Its
expression in terms of the fields of the D5-brane is\footnote{This expression
should be obtainable from the explicit form of the supercharges,
which one could get via Noether's theorem as in \cite{sato}.}
\be
Y^M=\int_\Sigma \vec{\Pi}\cdot {\narx}^M  \ ,
\ee
where $X^M$ is a vector in the transverse space to the D5 brane, \ie\
along the directions 1239. Thus
we recognize $Z_{el}$ as the central charge
$Y^{i\bar{\jmath}}$ appearing in the worldvolume
supersymmetry
algebra (in our expression \eqn{zel} the direction $M=9$ has been chosen
out of four equivalent ones).
The other charges occurring in the algebra are also interpreted
as intersections of branes along the lines of \cite{BGT, BPvdS}.

\section{The D5 in the D3 background: Hamiltonian analysis}
\label{ham.anal.}
In this section we will derive a BPS bound
on the energy of a D5-brane in the \bg\ geometry
of a stack of $N$
D3-branes, by showing that it is
bounded from below by a topological
quantity.
Let us thus begin by describing
the D3 background.
The ten-dimensional  metric is
\be
ds^2_{(10)} = H^{-1/2} d X_{\parallel}^2 + H^{1/2} \left( dr^2 + r^2
d \Omega_{(5)}^2 \right) \ . \ee Here $X_{\parallel}=(X^0, X^1, X^2,
X^3)$ are cartesian coordinates on $\bE^{(1,3)}$ and $d
\Omega_{(5)}^2$ is the line element on a unit five-sphere, which we
take to be parametrized by standard angular coordinates $\T, \Phi^i,
i=1, \ldots , 4$, where $\Phi^i$ are coordinates on a four-sphere.
Thus we have
\be
d \Omega_{(5)}^2 = d \T^2 + \sin^2 \T \, d \Omega_{(4)}^2 \ . \ee The
coordinate $r$ parametrizes the radial distance to the branes, so $r,
\T$, and $\Phi^i$ are spherical coordinates on the six-dimensional
space transverse to the branes. The harmonic function appearing above
is
\begin{equation}
H= a +\frac{ R^4}{  r^4}  \ .
\end{equation}
Here $a=0$ describes the near-horizon geometry $AdS_5 \times S^5$,
whereas \mbox{$a=1$} describes an asymptotically flat space-time. The
parameter $R$ measures the gravitational size of the stack of branes,
and in the case $a=0$ precisely coincides with the radius of both
$AdS_5$ and $S^5$. On top of the metric, the \su solution describing
the D3-branes is characterized by a non-vanishing Ramond-Ramond (RR)
5-form \fs \bea G^{(5)}&=& - H^{-2} \, H' \, dX^0 \wedge dX^1 \wedge
dX^2 \wedge dX^3 \wedge dr + 4 \, R^4 \, \omega_{(5)} \label{RR}~; \\
\omega_{(5)} &=& \sin^4 \T \, d\T \wedge \omega_{(4)} \ , \eea where
$\omega_{(n)}$ is the volume form on a unit $n$-sphere and $H' \equiv
\pa_r H$. Now let us consider a D5-brane coupled to the above
background. It will be described by the action
\be
S = - T_5 \int_{\Sigma} d^6 \sigma \, \sqrt{-\det (g+F)} +
T_5 \int_{\Sigma} \, A \wedge \gf    \ .
\label{fullaction}
\ee
Note that the RR field acts as source
for the \wv gauge field $A$ through
its coupling in the Wess-Zumino term
above. For all the configurations we
will consider, only the second term
in \eqn{RR} will contribute to this coupling.
The inclusion of this term is therefore
essential for the correct treatment
of such configurations.
\par
We are interested in passing to the Hamiltonian
formalism in order to derive a bound on the energy.
Let us first of all fix the static gauge by choosing
$\sigma=\{\t , \vp^i ; i=1, \ldots , 4\}$ as
coordinates on the D5-brane worldvolume and by
identifying
\be
X^0 = t \sac \T=\t \sac \Phi^i= \vp^i \ .
\ee
Since we will only be interested in configurations
which preserve the $SO(5)$ symmetry which rotates
the $\Phi^i= \vp^i$ coordinates, we also
restrict ourselves to
the following type of configurations:
\be
X^1 = X^2 = X^3 = 0 \sac r=r(t,\t) \sac A_t = A_t(t,\t)
\sac A_\t = A_\t(t,\t) \sac A_{\vp^i} = 0 \ .
\label{truncation2}
\ee
Under these conditions, the D5 action
reduces to
\be
S = T_5 \, V_{(4)} \, \int dt \, d\t \, \sin^4 \t \, \left[
- H \, r^4 \, \sqrt{r^2 + {r'}^2 - H \, {\dot{r}}^2 \,
r^2 - E^2} + 4 \, R^4 \, A_t \right] \ ,
\label{action}
\ee
where $\dot{r} \equiv \pa_t r \, , \, r' \equiv \pa_\t r \,
, \, E \equiv F_{0\t}$ and $V_{(4)}=8\,\pi^2/3$ is the volume
of a unit four-sphere. From now on we will set
$T_5 \, V_{(4)} \equiv 1$.
The canonical momenta $P$ and $\Pi$ conjugate
to the fields $r$ and $A_\t$ in the above action are
\bea
P &\equiv& \frac{\pa L}{\pa \dot{r}} \, = \,
\sin^4 \t \, \frac{H^2 \, r^6 \, \dot{r}}
{\sqrt{r^2 + {r'}^2 - H \, {\dot{r}}^2 \,
r^2 - E^2}} \nn
\Pi &\equiv& \frac{\pa L}{\pa \dot{A_\t}} \, = \,
\frac{\pa L}{\pa E} \, = \,
\sin^4 \t \, \frac{H \, r^4 \, E}
{\sqrt{r^2 + {r'}^2 - H \, {\dot{r}}^2 \,
r^2 - E^2}} \ .
\label{momenta}
\eea
The canonical phase-space Lagrangian density is defined by
\be
{\cal{L}} = \dot{A_\t} \, \Pi + \dot{r} \, P - L =
E\, \Pi + \dot{r} \, P - L - A_t \, \Pi'\ ,
\ee
where in the last step we have integrated $A_t' \, \Pi$ by
parts. Inverting the relations (\ref{momenta}) one can rewrite
$\cal{L}$ in the Hamiltonian form
\be
{\cal{L}}= \sqrt{\left( r^2 + {r'}^2 \right)
\left( \frac{P^2}{H \, r^2} + {\Pi}^2 + {\Delta}^2 \right)} +
A_t \, \left( - \Pi' - 4\, R^4 \, \sin^4\t \right) \ ,
\ee
where
\beq
\Delta \equiv H \, r^4 \, \sin^4\t~.
\eeq
The first term is just the desired energy density
\be
{\cal{H}}= \sqrt{\left( r^2 + {r'}^2 \right)
\left( \frac{P^2}{H \, r^2} + {\Pi}^2 + {\Delta}^2 \right)}\ ,
\ee
whereas the second term yields the `modified'  Gauss law
$\pa_\t \Pi = -4 \, R^4 \, \sin^4\t$. Remark that in the flat \bg
case we had as Gauss law $\vec{\nabla} \cdot \vec{\Pi}=0$, while here
we have a source term,
a direct consequence of the Wess-Zumino
coupling to $G^{(5)}$ in the D5-brane action. This is the difference
referred to in footnote~\ref{fn:Gauss}.
\par
The Gauss law constraint is solved
by \cite{Callan}
\be
\Pi(\nu,\t)= \frac{1}{2} \, R^4 \, \left[ 3 \, (\nu \, \pi - \t) +
3 \, \sin\t \, \cos\t + 2\,\sin^3 \t \, \cos\t \right]~~,
\label{GaussLawSol}\ee
where $\nu$ is an integration parameter. Its
meaning will become clear below (see also \cite{Callan}).
We thus use the solution of the Gauss law constraint, but do not use the
field equation for $r(\theta)$, which will rather be implied by BPS
conditions.
\par
We further consider static configurations.
The energy density reduces to
\be
{\cal{H}} = \sqrt{\left( r^2 + {r'}^2 \right)
\left( \Pi^2 + \Delta^2 \right)}  \ .
\ee
\par
The first key observation to prove the existence
of a BPS bound on the energy of the D5 is that
the above expression can be rewritten as
\be
{\cal{H}} = \sqrt{{{\cal{Z}}_{el}}^2 + r^2 \, \left(\Delta \,
\cos\t - \Pi\sin\t \right)^2 \, \left(\frac{r'}{r} - f \right)^2}
\label{right} \ee where \bea {\cal{Z}}_{el} &\equiv& r \,
\left(\Delta \, \cos\t - \Pi\sin\t \right) \, \left( 1 +
\frac{r'}{r}\, f \right)~~, \label{calZ} \\ f(a,\nu;r,\t)&\equiv&
\frac{\Delta (a,r,\theta) \, \sin\t + \Pi(\nu,\t)\, \cos\t}
{\Delta(a,r,\theta) \,\cos\t - \Pi(\nu,\t)\, \sin\t}~~. \eea The
function $r(\theta)$ has in general a limited range in $\theta$
from $\theta_i$ to $\theta_f$. The second key observation is
 that the integral
\be
Z_{el} \equiv \int_{\t_i}^{\t_f} d\t \, {\cal{Z}}_{el}
\label{Z}
\ee
is \emph{topological}, as we will show below.
This implies the desired BPS bound
on the D5 energy, namely
\be
E_{ren} + E_{gs}\equiv
\int_{\t_i}^{\t_f} d\t \, {\cal{H}} \geq |Z_{el}|~~,
\label{BPSbound}
\ee
through the following two inequalities
\be
E_{ren} + E_{gs} = \int_{\t_i}^{\t_f} d\t \, {\cal{H}} \geq
\int_{\t_i}^{\t_f} d\t \, |{\cal{Z}}_{el}| \geq |Z_{el}|\ .
\label{chain} \ee Above, $E_{gs}$ is the ground state or vacuum
energy of the D5-brane in the D3 background, and $E_{ren}$ is the
renormalized energy of a configuration, \ie\ the energy relative
to this ground state. Of course, $E_{gs}$ might be a divergent
quantity; only energies relative to it are physical. We will come
back to this issue below.
\par
The BPS bound \eqn{BPSbound}
is saturated if and only if both inequalities
above are. Saturation of the first one%
\footnote{If $r\left(\Delta \, \cos\t - \Pi\sin\t \right)=0$,
the introduction of $f$ in \eqn{right} is not allowed. In this case,
saturation of the BPS bound implies that the numerator of $f$ is zero
too, leading to
$\Pi(\nu,\theta)=\Delta(a,r,\theta)=0$, which implies $\theta=0$
or $\theta=\pi$, and
can thus be neglected.}
yields a first order differential equation
on the D5 embedding $r(\t)$, namely
\be
\frac{r'}{r} = f\ .
 \label{BPScondition}
 \ee
This equation is valid
for both the near-horizon case and the asymptotically flat one (note
that the function $f$ is different for the two cases). In the former
case, it was derived in \cite{Imamura} by imposing the preservation
of some fraction of global \wv supersymmetry. For the latter case, it
was proposed in \cite{Callan} as a plausible generalization of the
result in \cite{Imamura} and shown to imply the equations of motion
of the D5. In our approach, both cases can be dealt with at once, and
in both of them the BPS equation emerges from a bound on the D5
energy.
\par
On top of \eqn{BPScondition}, we still have to impose
saturation of the second inequality in \eqn{chain}
to saturate the energy bound \eqn{BPSbound}.
Such inequality is saturated when
${\cal{Z}}_{el}$ does not change sign
in the integration region $[\t_i,\t_f]$.
This condition determines
the possible values of
$\t_i$ and $\t_f$ (for BPS solutions), where boundary
conditions for $r$ should be supplemented.
When \eqn{BPScondition} holds, the sign
of ${\cal{Z}}_{el}$ is determined
by the sign of
\be
\Delta \cos\t-\Pi\sin\t = a\, r^4\, \sin^4\t\,\cos\t +
\frac{3}{2}\, R^4 \, \sin\t\, \eta(\t)~~,
\label{signfactor}
\ee
where the function
\be
\eta(\t) \equiv \t - \pi \, \nu - \sin\t\,\cos\t
\ee
is monotonically increasing for $\t\in[0,\pi]$.%
\footnote{Recall that $0\leq\t_i\leq\t_f\leq\pi$
since we identified $\t$ with $\T$,
one of the angular coordinates on the $S^5$.}
In \cite{Callan} the restriction was made to $\nu\in [0,1]$.
We make the same restriction in the main text.%
\footnote{If one were to consider solutions with $\nu$ outside
this range, $\eta$ would have a definite sign, so that, for $a=0$,
the
solution $r(\t)$ can extend over the whole interval $[0,\pi]$,
developing spikes both at $\t=0$ and $\t=\pi$.}
In this case, $\eta$ changes sign precisely once in this
interval at $\t_0$ such that $\eta(\t_0)=0$.
Therefore, in the case $a=0$, the second inequality
in \eqn{chain} is saturated if
\be
\t_i=0\sac \t_f=\t_0
\label{range0i}
\ee
or
\be
\t_i=\t_0\sac \t_f=\pi~~.
\label{range0ii}
\ee
These two possibilities
coincide with those found in \cite{Callan}, where
they were determined by explicitly
solving the BPS equation \eqn{BPScondition}.
\par
Although the procedure is conceptually the
same, in the case $a=1$ we cannot determine
from this analysis
the possible range of $\t$.
To determine the sign of
\eqn{signfactor} we would need the
explicit form of the solution $r(\t)$.
These solutions were studied
in \cite{Callan} only numerically (in the case
$a=1$): no analytic solution was found.

Let us come back to the topological nature of
$Z_{el}$.
Since the density ${\cal{Z}}_{el}$ can be rewritten as a total derivative,%
\footnote{Only the Gauss law solution \eqn{GaussLawSol} has to be used
to deduce \eqn{TotDeriv}. Neither the field equation for
$r(\t)$ nor the BPS condition \eqn{BPScondition} are needed.}
\be
r\,(\Delta
\cos\t-\Pi\sin\t)\left(1+\frac{r'}{r}\,f\right)=
\frac{d}{d\t}\,\left\{\Pi\,r\cos\t+
\left(\frac{a}{5}+\frac{R^4}{r^4}\right)(r\sin\t)^5\right\}~~,
\label{TotDeriv}\ee
it follows that
\be
Z_{el} =\left[\Pi\,r\cos\t\right]_{\t_i}^{\t_f}
+\left[\left(\frac{a}{5}+\frac{R^4}{r^4}\right)
(r\sin\t)^5\right]_{\t_i}^{\t_f}
\label{Zintegrated}
\ee
depends only on the boundary conditions
$r(\t_i)$ and $r(\t_f)$. This means that,
for fixed values of $r(\t_i)$ and $r(\t_f)$,
$Z_{el}$ is invariant under
local variations of the fields.
It is in this sense that it
is a topological quantity.
Configurations saturating the bound
(\ref{BPSbound}) minimize the energy
for fixed boundary conditions,
and therefore they automatically
solve the equations of motion.
\par
For $a=1$,
the second term
is precisely the ground state energy
of the D5. Indeed,
consider an infinite
planar D5-brane intersecting
the stack of D3-branes at a point
\be
\ba{ccccccccccl}
D3: &1&2&3&\_&\_&\_&\_&\_&\_  &\quad \mbox{background}   \nn
D5: &\_&\_&\_&4&5&6&7&8&\_ &\quad \mbox{worldvolume.}
\ea
\label{D5gs}
\ee
The D5 worldspace $\Sigma$ spans a five-dimensional
hyperplane in the transverse space to the
D3-branes, which we can take as the $\theta=\pi/2$
hyperplane. The energy of such a configuration is
\be
E_{gs}=\int_\Sigma d^5 \sigma\,\sqrt{-\det g}=
\int_0^{r_f} d r \, r^4\,H(r)=
\left[\left(\frac{a}{5}+\frac{R^4}{r^4}\right)\, r^5\right]_
0^{r_f}~~,
\ee
which can be seen to agree with the second
term in \eqn{Zintegrated} by making use
of the form of the solutions found in \cite{Callan}.
\par
In the case $a=0$, the second term in
\eqn{Zintegrated} vanishes for all the
solutions of \eqn{BPScondition} (again, see
\cite{Callan}). We will see below that
the first term in \eqn{Zintegrated}
has a clear physical interpretation
not associated with any ground state energy,
so we conclude that the D5 ground state for $a=0$
is such that its energy vanishes.
We interpret this fact in the following
way. Let us understand
the ground state of the D5 in the near
horizon geometry of the D3-branes as a
limit of the D5 ground state in the full D3
geometry. The latter is represented
by the array \eqn{D5gs}. One can think of it
as an intersection of orthogonal D5 and
D3 branes in a point, but this
is not necessarily the case: since it is
a BPS configuration, both kinds of branes can
be separated along the 9th direction while preserving
the BPS character. Hence, generically,
in the full D3 geometry, the D5 in its
ground state does not intersect the D3-branes.
Taking the near horizon limit means zooming up
the region close to the latter.  If the D5
was placed at a certain finite distance along
$X^9$, then it will `disappear to infinity' when one
focuses on regions arbitrarily close to
the D3-branes. In this sense, its
ground state energy in the full geometry
will not show up after taking the near
horizon limit.
\par
The D5 ground state
we end up with taking this limit can be identified
with the $\nu=1$ configuration found in \cite{Callan}.
The reason is that configurations with $\nu\neq 1$ in
\cite{Callan} represent a D5-brane infinitely
far away from the D3-branes, connected to them by
$(1-\nu)\,N$ strings. The limit $\nu \rightarrow 1$
can thus be interpreted as the limit in
which only the D5 in its ground state `at infinity' remains.
This remark is important because although
in the strict limit the D5 disappears,
in taking such a limit there is always a
selected direction, namely $X^9$.
This will be essential in the next section.
\par
Having identified $E_{gs}$ for both the $a=0$
and the $a=1$ cases,
we can rewrite the bound \eqn{BPSbound} as
\be
E_{ren} \geq |Z_9| \sac Z_9 \equiv
\left[\Pi\,r\cos\t\right]_{\t_i}^{\t_f}
\label{renbound}
\ee
The physical meaning of $Z_9$ can (again) be understood
by evaluating it for the solutions of
\eqn{BPScondition}. Restoring the factor $T_{5} \, V_{(4)}$
the result is%
\footnote{In the case $a=0,\nu=1$ (see \cite{Callan})
there is an additional contribution to $Z_9$
given by $-\Pi(\t_i) \, r(\t_i)$. We neglect it because
it is finite, whereas $Z_9$ is divergent. It is precisely
the form of this divergence that we would like
to interpret.}
\be
Z_{9}=n \, T_f \, L\,, \qquad n=\left\{
\begin{array}{ll}
(1-\nu) \, N & \mbox{for $a=0$} \nn
(\ft12-\nu) \, N & \mbox{for $a=1$}
\end{array}
\right.
\label{Zel}
\ee
where $L\equiv r(\t_f)$ for $a=0$ and
$L\equiv r(\t_f) \cos(\t_f)$ for $a=1$.
The above expression clarifies the physical interpretation
of $Z_{9}$. It is the energy of $n$
fundamental strings of length $L$ and
tension $T_f$ joining together the D5 and the D3-branes.
{}From the ten-dimensional space-time point of view
this corresponds to the triple intersection
\be
\ba{ccccccccccl}
D3: &1&2&3&\_&\_&\_&\_&\_&\_ &\quad \mbox{background}    \nn
D5: &\_&\_&\_&4&5&6&7&8&\_ &\quad \mbox{worldvolume}     \nn
F1: &\_&\_&\_&\_&\_&\_&\_&\_&9 &\quad \mbox{BPS solution.}
\ea
\label{triple}
\ee
In our analysis, the D3 is
the background and the F1
plays the role of a soliton
of the D5 worldvolume theory.%
\footnote{Triple intersection from the point
of view of worldvolume theories have been studied in \cite {BGT}, see also
\cite{glw}.}
\par
We have chosen the name $Z_9$
because of the direction in which the string
stretches. In the case $a=0$, the
D5 is infinitely far away from the D3-branes,
so $L$ is divergent. In the case $a=1$,
the D5 is asymptotically a flat
hyperplane
and $L$ coincides
with its distance
to the point $r=0$, where the D3-branes are.
\par
The solutions of the BPS equation (\ref{BPScondition}) are thus the
D5 \wv description (spike-like solutions) of the configuration
\eqn{triple}. In the next section we will provide further evidence in
favor of the interpretation of $Z_{9}$ as a charge associated to the
ending of fundamental strings on the D5 \wv from its appearance as a
central charge in the D5 \wv supersymmetry algebra.

\section{Algebras}
\label{algebras}
In the previous section we have understood the BPS equation
\eqn{BPScondition} from a
bosonic point of view.
Now we will show that the charge $Z_{9}$,
which sets the bound \eqn{renbound}, is
a central charge in the worldvolume
supersymmetry algebra of the D5 in the D3
background. To do so, we will
determine this algebra. The cases $a=0$ and $a=1$
will be treated separately, since the results
are different. However, the procedure
to determine the algebra is conceptually
the same, and in principle applies to any
brane in any curved background.
\par
The starting point is to identify the ground state
configuration of the brane in the given
background. Two remarks are in order
here. First, one should
choose it according to which
excited configurations one would like to
study. From a space-time point of view,
this corresponds to which type of
triple intersections one would like to consider.
For instance, suppose instead of \eqn{triple}
we had been interested
in studying the intersection
\be
\ba{ccccccccccl}
D3: &1&2&3&\_&\_&\_&\_&\_&\_&\quad \mbox{background}    \nn
D5: &\_&2&3&4&5&6&\_&\_&\_ &\quad \mbox{worldvolume}      \nn
D3: &1&\_&\_&4&5&\_&\_&\_&\_&\quad \mbox{BPS solution.}
\label{triple1}
\ea
\ee
Then the vacuum configuration would have consisted of
an static infinite planar D5-brane intersecting the
stack of D3-branes in a 2-brane. The BPS-solution would have
been described by the scalar $X^1$ and an excited BI gauge field.
\par
Second, the starting point is purely bosonic, since one need only
consider a bosonic background.
\par
Next, one determines which isometries
of the background preserve the ground
state configuration; these isometries
give rise to linearly realized symmetries
of the brane worldvolume theory \cite{conffadS}: they
constitute the worldvolume bosonic symmetry
algebra. Part of them are associated
to worldvolume coordinate transformations and part of them
to R-symmetry.
\par
To determine the whole
worldvolume supersymmetry
algebra, one first imposes that the supersymmetry
generators are in the (fundamental) spinorial
representation of the algebra of worldvolume coordinate transformations.
Moreover, they should be in a representation of the R-symmetry algebra.
That is usually sufficient to \emph{fix} the number
of supercharges of the worldvolume supersymmetry algebra.
In the case $a=0$ one can make use of the classification of Nahm
\cite{Nahm} of super-adS or superconformal algebras%
\footnote{In principle, the supersymmetry generators are determined from
the
Killing spinors of the above-mentioned bosonic ground state configuration.
One could have started
from the whole isometry superalgebra of the (super)background
and have determined which of these superisometries preserve the
ground state of the brane.\label{background}}.

\par
To definitely fix this algebra,
only the anticommutator of these
supercharges needs still to be determined.
This is done by allowing all the \emph{original} bosonic
generators not forbidden by Jacobi identities
to occur.
\par
The algebra constructed this way, is what we
call the \emph{worldvolume supersymmetry algebra}.
The only bosonic generators it contains,
are those originating from bosonic isometries
of the background.
\par
The final step in the procedure is
to enlarge this algebra along the
lines of \cite{BGT} to what we will call
the \emph{maximally extended worldvolume supersymmetry algebra}.
This is done by constructing the most
general anticommutator of the supersymmetry
generators allowed by symmetry by
adding all possible central
extensions (see footnote~\ref{fn:central}). The algebra obtained in this
way is the one which encodes all the BPS
intersections. Such intersections involve the brane
whose worldvolume theory we are considering,
the branes creating the background and a third
one. The latter is the one associated to
the central charges we have enlarged
the worldvolume algebra with.
\par
Note that this procedure does not completely fix
the superalgebra yet, because so far one has
not determined the commutators of the central
charges with the supersymmetry
generators, with the original bosonic generators
and with themselves! An explicit computation
of the charges via Noether theorem would
allow one to fix this ambiguity. However, consistency with Jacobi
identities fixes in many cases this algebra, as we will see below.
\par
In the rest of this section we will
illustrate the above general procedure
by applying it to a D5 brane
in a D3 background. We will show
the appearance of $Z_9$, and also interpret
the central charges
in the supersymmetry algebra.

\subsection{The near-horizon algebra}
The first task is to determine
which bosonic isometries of the background
preserve the ground state of the D5-brane.
Recall that the near-horizon
D3 background has as worldvolume superalgebra $SU(2,2|4)$, with bosonic isometry group
\mbox{$SO(4,2)\times SO(6)$}.
As already discussed in
section \ref{ham.anal.}, when the D5 is in its ground state
its worldspace  consists of a
hyperplane extending along directions 45678, located
at arbitrary constant values of $X^i$, $i=1,2,3$,
for instance $X^i=0$,
and at \mbox{$X^9 \rightarrow \infty$}.
\par
This configuration is preserved by the subgroup
\mbox{$SO(2,1)\times SO(5)\times SO(3)$} of the
background bosonic isometries.
Indeed, the condition \mbox{$X^1=X^2=X^3=0$} is preserved by
an \mbox{$SO(2,1)\times SO(3)$} subgroup of the $SO(4,2)$
factor, and selecting the directions 45678
(or alternatively, the 9th) breaks the $SO(6)$ factor to
$SO(5)$.
\par
The factors of the symmetry group have the following
interpretation: the $SO(2,1)$ contains time translations,
dilations and boosts for the radial coordinate
$r=\sqrt{\sum_{i=4}^{9}(X^i)^2}$.
 The $SO(5)$
contains the D5 worldspace rotations and the $SO(3)$ is the group of
D3 worldspace rotations, which is, from the point of view of the D5
brane supersymmetry group, the R-symmetry.
\par
We have thus identified the bosonic worldvolume
symmetry group. The 32 fermionic generators of the D3 worldvolume
algebra are in the representation $(4,\bar{4})+(\bar 4,4)$ of
$SO(4,2)\oplus SO(6)$. The intersection with $D5$ preserves half of
them, which are still spinors of $SO(2,1)$, $SO(5)$ as well as
$SO(3)$.
 They are thus in the $(2,4,2)$ representation of the bosonic
worldvolume algebra. Using the classification of Nahm \cite{Nahm}
(for the terminology we use, see the review in \cite{Triest98}) we
can identify therefore the {\em worldvolume superalgebra as $OSp(4^*|4)$}, using
the identification $SO^*(4)=SO(2,1)\times SO(3)$, and $USp(4)=SO(5)$.
The superalgebra is in this way completely fixed. On the worldvolume
half of the supersymmetries are ordinary $Q$-supersymmetries, and the
other half are $S$-supersymmetries, as distinguished by the
eigenvalue of the dilatations. This conformal worldvolume superalgebra originates
from the super-adS symmetry in spacetime \cite{conffadS,Triest98}. In
the latter form the fermionic generators $Q$ and $S$ are unified in a
supersymmetry $Q$. In this formulation the
anticommutators of these supersymmetries are
\begin{equation}
\left\{Q_\alpha ^i,Q_\beta^j\right\}=C_{\alpha
\beta}R^{[ij]}-\ft18
\gamma^{\mu\nu} _{\alpha \beta}M_{[\mu \nu]} \delta^{ij}\ ,
\end{equation}
where $\alpha ,\beta=1,\ldots 4$ and $\mu ,\nu=4,\ldots 8$ are
spinor and vector indices of $SO(5)$, respectively. The
$i,j=1,\ldots 4$ are vector indices of $SO^*(4)$. At the right
hand side appear $R^{[ij]}$, the generators of $SO^*(4)$, and
$M_{[\mu \nu]}$, the generators of $SO(5)$. In the conformal
notation of this algebra, the latter appear only in the $\{Q,S\}$
anticommutator. The commutators of the bosonic algebra and the
supersymmetries just reflect that the supersymmetries are in the
$(2,4,2)$ representation.
\par
Having identified the worldvolume algebra as $OSp(4^*|4)$, we now
look for the extended worldvolume algebra.
In the spirit of \cite{BGT} we thus enlarge this algebra by all possible
generators that can appear on the right hand side
\begin{equation}
\left\{Q_\alpha ^i,Q_\beta^j\right\}=C_{\alpha
\beta}R^{[ij]}-\ft18
\gamma^{\mu\nu} _{\alpha \beta}M_{[\mu \nu]} \delta^{ij}
+\gamma^\mu_{\alpha \beta} U_\mu^{[ij]}
+ \gamma^{\mu\nu} _{\alpha \beta}Z_{[\mu \nu]}^{(ij)_s} \ ,
\label{QQa0E}
\end{equation}
where $(ij)_s$ indicates that $Z$ is symmetric traceless. The Jacobi
identities do not allow that the generators $U$ and $Z$ commute with the
supersymmetries. Indeed this is clear {\it e.g.} from the Jacobi identity
$[Q,Q,Z]$, due to the fact that $SO^*(4)$ and $SO(5)$, which rotate the supersymmetries, appear
in the right hand side of \eqn{QQa0E}. All the bosonic generators
should thus be part of this simple Lie superalgebra which contains the
conformal algebra. Looking through the list of all superalgebras \cite{LieSA}
(see \cite{Triest98} for a convenient table), one finds that the
{\em extended worldvolume superalgebra is $OSp(1|16)$}. The right hand side
of \eqn{QQa0E} is a symmetric $16\times 16$ matrix $M_{AB}$ where $A=(\alpha
i)$. Introducing the symplectic metric $\Omega^{AB}={\cal C}^{\alpha\beta}
\delta^{ij}$, one obtains the algebra
\begin{eqnarray}\label{commMQ}
&&\left\{ Q_A,Q_B\right\}=M_{AB}\sac
  \left[M_{AB},Q_C\right]= Q_{(A}\Omega_{B)C}\nonumber\\
&&  \left[ M_{AB},M_{CD}\right] = \Omega_{A(C}M_{D)B}+ \Omega_{B(C}M_{D)A}\ ,
\end{eqnarray}
which thus defines the full algebra. The $Sp(16)$ generators are
decomposed in \eqn{QQa0E} under
$SO(2,1)\times SO(5) \times SO(3)$  as
\begin{equation}
[(3,1,1)+(1,1,3)]+(1,10,1)+[(3,5,1)+(1,5,3)]+(3,10,3)\ .
\label{QQa0Erep}
\end{equation}
\par
For the splitting of fermionic generators in $Q$ and $S$ supersymmetries,
we have to write the $SO^*(4)$ vector index $i$ as $(\kappa \pm)$,
where $\kappa=1,2$ is a spinor index of $SO(3)$ and $\pm$ indicates the
$SO(2,1)=SU(1,1)$ index. The ordinary ($Q$) supersymmetries are those where this
latter index is $+$. The anticommutator of these
ordinary supersymmetries is then obtained from \eqn{QQa0E} as
\begin{equation}
\label{QQQa0E} \left\{Q_\alpha
^\kappa,Q_\beta^\lambda\right\}=C_{\alpha \beta}\tilde{R}
\epsilon^{\kappa\lambda} +\gamma^\mu_{\alpha \beta} \tilde
U_\mu\epsilon^{\kappa\lambda} + \gamma^{\mu\nu} _{\alpha
\beta}\tilde Z_{[\mu \nu]}^{(\kappa\lambda)}\ ,
\end{equation}
where $\tilde R$, $\tilde U$ and $\tilde Z$ are the $(++)$ parts of $R$, $U$ and $Z$.
\par
The charge $Z_{9}$ is a component of a 3-vector of
$SO(2,1)$. Indeed, it is proportional to the value of $r$ at the
singularity, see \eqn{Zel}, and $(r,rt, -rt^2+R^2/r)$ is a 3-vector,
the one used to embed $AdS_2$ in a 3-dimensional flat
space (see e.g. (17) in \cite{Triest98} with $z=r^{-1}$ and $x^\mu=t$).
Under the remaining factors of the bosonic symmetry group it
is invariant, and thus we conclude that $Z_{9}$ is part of a
$(3,1,1)$. It can thus be combined with $(E,D,K)$, the generators of
energy, dilatations and Lorentz boosts in $(t,r)$, to appear
in $R$. The $(3,1,1)$
part of the generators $R^{[ij]}$ in \eqn{QQa0E} is the triplet
of $SO(2,1)$ whose $(++)$ component is%
\footnote{In the IIB susy algebra in 10 dimensions, the central
charge $Z_9$ and the energy $E$ occur separated. The breakdown of
this algebra to the $Q$ supersymmetry part of the extended
worldvolume algebra \eqn {QQQa0E} should lead to this combination.}
\begin{equation}\label{combine}
\tilde R=(E-Z_{9})\ .
\end{equation}
(As before,
for the definition of $E$
we have put the energy of the vacuum configuration of the
D5 equal to zero.)
The fact that this combination $E-Z_{9}$ appears on the right-hand side of the
anticommutator of the supersymmetry generators implies
that a configuration with non-vanishing string charge along the
9-direction (and the other central charges equal to zero) preserves
all 16 supersymmetries if the energy is chosen to saturate the
corresponding BPS-bound.

Indeed, from a space-time point of view it
is easy to see that, having performed the projections onto
supercharges preserving the D3-D5 intersecting (or overlapping) on a
point, one can put an additional string in the remaining direction
`for free' (see, for instance, \cite{Callan}).
\par
To summarize the main point of this subsection, we have shown that
$Z_9$ has indeed the right quantum numbers to be interpreted as a
central charge in the worldvolume supersymmetry algebra of
the D5-brane in the near-horizon limit of the D3 background;
and that doing so makes clear that the
corresponding BPS-solutions (the strings in
the 9th-direction) preserve all 16 supersymmetries of
this algebra.

\subsection{The non-near-horizon case}
We now move on to discuss the $a=1$ case. The background
bosonic isometry group is $ISO(3,1)\times SO(6)$.
The vacuum configuration consists of a hyperplane
extending along directions 45678, located
at arbitrary values of $X^i$, $i=1,2,3$ and $X^9$.
Therefore, the worldvolume bosonic symmetry
group is $\bR \times SO(5)\times SO(3)$.
Both the worldvolume superalgebra and its maximal
extension are obtained by applying the general procedure, as was done
in the near-horizon case. However, it can also be obtained
from the $a=0$ results. We find this alternative illustrative,
so we describe it here.
\par
Comparing to the $a=0$ case,
we see that the $SO(2,1)$ conformal symmetry appearing there, has been broken
to just time translations \bR, its \Poin\ part.
In general, the \Poin\ algebra is not
a simple algebra, it is the semi-direct sum of translations and the
Lorentz algebra. In this case, the latter is trivial, because there
is just the time direction\footnote{Note that not the full worldvolume
coordinate transformation algebra is conformal for $a=0$, but only
the part corresponding to the intersection point, which is the 1-dimensional
conformal algebra $SO(2,1)$. The worldvolume coordinate
transformations contain also the $SO(5)$ rotations, which are from
the point of view of the superconformal theory of the point, an
R-symmetry.}.

Similarly, the worldvolume
superalgebra $OSp(4^*|4)$ from the $a=0$ case
is broken, for $a=1$, to an algebra which is the super-\Poin\
subalgebra of the 1-dimensional superconformal algebra%
\footnote{This super-\Poin\ algebra can alternatively be seen as a contraction of the
super-adS algebra. We come back to these issues in
appendix~\ref{app:superalgebras}.}. The latter
contains as first part the $Q$-supersymmetries and what appears in
their anticommutator, \ie\ the (time) translations. This part satisfies
Jacobi identities by itself. But it is not a simple superalgebra.
The full super-\Poin\ algebra is the semi-direct sum of this first part
with the Lorentz algebra, and the
automorphism group of the supersymmetries. The conformal algebra
is here $SO(2,1)$, and has no Lorentz subalgebra. The automorphism
group is $SO(3)\oplus SO(5)$. The $SO(5)$ rotations belong to the worldvolume
algebra for the D5, being its worldspace rotations. For D5,
$SO(3)$ is the R-symmetry. This is the worldvolume
algebra for $a=1$.
\par
The extended worldvolume superalgebra can again be obtained from the one in
the near-horizon case using the breakdown of the conformal algebra.
With the breaking of the conformal algebra, also its representations
are cut. The $Q$ and $S$ supersymmetries together form a
representation of the conformal algebra, here $SO(2,1)$.
The supersymmetries are distinguished according
to their eigenvalue under the dilatations, which is, with suitable
normalization, $\pm\ft12$. The ones with positive weight, \ie\ the $Q$
supersymmetries, are present, while the ones with negative
eigenvalue, the $S$ supersymmetries, are not any more present.
Similarly the central charges $U$ and $Z$ in (\ref{QQa0E}) can be
split according to their eigenvalue under the dilatation operator,
which is 1,0 or $-1$. Only the former appear in the anticommutator of
two $Q$ supersymmetries, and they thus remain in the super-\Poin\
algebra, while the others do not. Those which remain are the $\tilde
U$ and $\tilde Z$ as in (\ref{QQQa0E}). The bosonic generators in
there mutually commute and commute with the supersymmetries. The full
extended worldvolume algebra thus consists of these generators and in
general the Lorentz generators (not present in this case) and the
R-symmetry generators of the superconformal algebra, which are here the $SO(3)$ part of the
$SO^*(4)$ generators $R^{[ij]}$ as R-symmetries for the D5 brane, and $M_{\mu\nu}$, the generators of
$SO(5)$ worldspace rotations.
\par
Still a third possibility to obtain the
full extended worldvolume superalgebra
consists in starting from the
one of a D5-brane in a flat background
\cite{BPvdS} and restricting to the generators preserved when
a D3-brane is added to the configuration.
\par
The result can be described as a contraction of
$OSp(1|8)$. The latter contains the anticommutator (\ref{QQQa0E}),
but with bosonic operators that do not commute mutually or with the
supersymmetries, similar to (\ref{commMQ}). The contraction consists
in scaling the fermionic generators with $x$, the bosonic ones with
$x^2$, and then taking the limit $x\rightarrow \infty$.
\par
As in the $a=0$ case we have the identification (\ref{combine}).
Note that now $Z_9$ is a singlet under the whole
symmetry group, as corresponds to the fact that
now, in contrast to the $a=0$ case,
this group does not act on the coordinate $r$
to which $Z_9$ is proportional.
Apart from  this difference, the
conclusions about $Z_9$ are the same as in the
$a=0$ case: it has the interpretation of a fundamental
string in the 9th direction%
\footnote{Note that from the symmetry group point of view, $Z_9$,
being a singlet, could also be interpreted as a 0-brane inside the
D5; however, since there is no 0-brane in the IIB theory, we conclude
that $Z_9$ must be associated to a fundamental string in the 9th
direction. (A D-string would not yield a BPS configuration.)}
and this string breaks no additional supersymmetries if
the energy is properly chosen.

\subsection{Interpretation of the central charges}
Let us start with the $a=1$ case.
In addition to the combination (\ref{combine}) of energy minus string charge, the
algebra \eqn{QQQa0E} contains the charges $\tilde U_\mu$ and $\tilde Z_{[\mu\nu]}^{
(\kappa\lambda)}$. Using standard arguments (see for instance
\cite{townsend} for examples in the M-theory context), one can show
that these correspond to 1/2 BPS solutions of the worldvolume theory.
Bearing in mind the interpretation of the R-symmetry as rotations in
the D3-brane worldspace, these charges suggest a space-time
interpretation of the corresponding worldvolume solutions
\cite{BGT}. In our case they can be interpreted as a third brane
intersecting the D5-brane in the D3 background, preserving 1/8
supersymmetry (4 supercharges). We now proceed to show that this
interpretation is indeed correct. (For an analogous discussion for a
D5-brane in a flat background, see \cite{BPvdS}.)
\par
The charge $\tilde U_\mu$ transforms in the (5, 1) representation of
$SO(5)\times SO(3)$. The BPS solution corresponding to it can be
interpreted in space-time as a triple intersection of a D5-brane,
a D3-brane and either
\begin{itemize}
\item an NS5-brane in {\it e.g.} the 45679 directions,
\item a D1-brane in {\it e.g.} the 4 direction, or
\item a D7-brane in  {\it e.g.} the 1234567 directions.
\end{itemize}
These configurations indeed preserve 1/8 supersymmetry.
Let us {\it e.g.} illustrate the first one.
A charge in the (5,1) representation
of $SO(5)\times SO(3)$ selects either 1 or (by hodge
duality) 4 of the 5 worldspace directions 45678, and either
none or 3 of the 123 directions. Suppose we take it to select
4 worldspace directions, for instance 4567, and none among the
123. Then we would have to associate it to a 4-brane inside
the D5-brane, but there is no 4-brane in the IIB theory.
However, since the 9th
direction is inert under the worldvolume symmetry group,
we can assume that we have a 5-brane instead,
extending along directions 4567 \emph{and 9}. Hence,
we end up with the interpretation  of an NS5-brane along directions
45679 (a D5-brane along these directions would not be BPS).
\par
Analogously, $\tilde Z_{[\mu\nu]}^{ (\kappa\lambda)}$ corresponds to one of the
following branes:
\begin{itemize}
\item a D5-brane in {\it e.g.} the 23456 directions,
\item a D3-brane in {\it e.g.} the 145 directions, or
\item an NS5-brane in  {\it e.g.} the 23459 directions.
\end{itemize}
\par
So far, we have interpreted the  charges for the $a=1$ case. In the
$a=0$ case, the same charges appear in the $\{Q, Q\}$ anticommutator,
and there are many more in $\{Q, S\}$ and $\{S, S\}$. These do not
represent other solutions. Indeed, as explained in
the previous subsections, and as will be illustrated in
appendix~\ref{app:superalgebras}, these charges are partners of those
in $\{Q, Q\}$ in representations of the bosonic conformal algebra.
This is similar to the special conformal transformations $K$, which
are just partners of the translations $P$.
In fact, the charges in the $\{Q, Q\}$ anticommutator constitute a maximal
commuting subset.
\section{Conclusions}     \label{ss:conclusions}
In this paper we have extended the BPS method for branes propagating
in a flat background to the case of branes in a non-trivial,
curved background. This method has two complementary aspects.
\par
On the one hand, the energy density is written as a sum of squares in a
handy
way. This allows the derivation of a BPS bound and
the corresponding first order
BPS equation.
\par
On the other hand, the BPS bound is interpreted in terms of a
central charge appearing in the maximally extended
worldvolume supersymmetry algebra. This algebra is constructed by
first analyzing the bosonic symmetries preserved by the vacuum
solution and second constructing the most general anticommutator of
the corresponding supersymmetry generators, including all possible
central charges.
\par
We have explicitly analyzed the case of a D5-brane in the background
of a D3-brane, both in the near and non-near horizon cases. We have
obtained in a unified way the BPS equations previously found in
\cite{Imamura} and \cite{Callan}.
\par
In the case of the near horizon geometry, the bosonic symmetries of the
vacuum
solution
are $SO(2,1)\times SO(5)\times SO(3)$, the worldvolume algebra is
$OSp(4^*|4)$, and the corresponding
maximally extended worldvolume superalgebra
is
$OSp(1|16)$, whose algebra
contains $16$ supercharges. The BPS bound is associated to a central
charge,
representing strings ending on the D5-worldvolume. Thus the existence of a
BPS
configuration preserving all $16$ supercharges has been understood at the
level of the worldvolume supersymmetry algebra.
\par
For the   non-near horizon case, the bosonic symmetries of the vacuum
solution
are $\Rbar\times SO(5)\times SO(3)$ and the maximally extended
worldvolume supersymmetry algebra
is a  contraction of $OSp(1|8)$, which contains $8$ supercharges.
The charges appearing in
the worldvolume algebra are interpreted in terms of multiple
intersections of
branes.

\medskip
\section*{Acknowledgments}
\noindent
We would like to thank Piet Claus for interesting discussions.
B.C. and A.V.P. thank the Physics Department of the University of
Barcelona for
the hospitality during a fruitful visit in which a substantial part
of this work was
performed. D.M. is supported by a fellowship from
Comissionat per a Universitats i Recerca de la Generalitat de
Catalunya. This work was supported in part by
the European Commission TMR programme ERBFMRX-CT96-0045,
AEN98-0431 (CICYT), GC 1998SGR (CIRIT),
 NSF Grant PHY9511632 and the Robert A. Welch Foundation.
\newpage
\appendix
\section{Notations}    \label{app:notations}
We use $\alpha ,\beta, ...$ for spinor indices. Gamma matrices are
then of the form $\left( \gamma_\mu \right) _\alpha {}^\beta$, but
indices may be lowered or raised using the charge conjugation matrix
${\cal C}^{\alpha \beta}$ and its inverse, which has lower indices:
${\cal C}^{\alpha \beta}{\cal C}_{\gamma\beta}=\delta^\alpha
_\gamma$. This allows us to use NW-SE conventions for raising and
lowering indices, {\it e.g.}: $\lambda^\alpha ={\cal C}^{\alpha
\beta}\lambda_\beta$ and $\lambda_\alpha =\lambda^\beta{\cal
C}_{\beta\alpha }$.
When we write in the text $\gamma^\mu _{\alpha \beta}$, this is thus
$\gamma^\mu _{\alpha}{}^\gamma{\cal C}_{\gamma \beta}$.
\section{Superalgebras and subalgebras}\label{app:superalgebras}
In the main text we referred to various \Poin, adS and conformal
superalgebras, and their relations as subalgebras or contractions.
Here we will repeat the standard relations and generalize them to
the extended worldvolume algebras. To have a less trivial example
than $SO(2,1)$ as in the main text, we treat the algebra of the
D3 background, which is ${\cal N}=4$ supersymmetric in $d=4$.
\par
The ordinary supersymmetries in that case are the 16 generators
$Q_\alpha ^i$ where $\alpha $ denotes the spinor index of $SO(3,1)$,
and $i=1,\ldots 4$ is a vector index for $SU(4)$.
We use chiral notations, \ie
\begin{equation}\label{chiralQ}
Q_\alpha^i=\ft12(1-\gamma_5)_\alpha{}^\beta Q_\beta^i\sac
Q_{\alpha i}=\ft12(1+\gamma_5)_\alpha{}^\beta Q_{\beta i}\ ,
\end{equation}
such that complex conjugation can be done by raising or lowering the index $i$
(see e.g. appendix A in \cite{trsummer} for more explanations).

The super-\Poin\
algebra has non-zero (anti)commutators
\begin{eqnarray}
\left\{Q_\alpha ^i,Q_{\beta j}\right\}&=&\gamma^\mu _{\alpha \beta}P_\mu
\delta^i_j \nonumber\\
\left[M_{\mu\nu},M_{\rho\sigma}\right]
&=&\eta_{\mu[\rho}M_{\sigma]\nu}
-\eta_{\nu[\rho}M_{\sigma]\mu}\nonumber\\
\left[P_\mu , M_{\nu\rho}\right]&=&\eta_{\mu[\nu}P_{\rho]}\sac
\left[ M_{\mu\nu} , Q_{\alpha}^i \right] = -\ft14
(\gamma_{\mu\nu})_{\alpha}{}^\beta Q^i_\beta
\nonumber\\
\left[ U^i{}_j,Q^k\right]&=& \delta^k_j Q^i\sac
\left[U^i{}_j,U^k{}_\ell\right]=\delta^k_j U^i{}_\ell -\delta^i_\ell
U^k{}_j\ ,\label{superPoincare}
\end{eqnarray}
where the latter is the $U(4)$ automorphism group. Again on $U$
complex conjugation is done by raising or lowering indices, and
the antihermiticity of the generators is then reflected in
\begin{equation}\label{Uunitary}
(U_A)_i{}^j\equiv\left((U_A)^i{}_j \right) ^* =-(U_A)^j{}_i\ .
\end{equation}
The super-\Poin\ algebra is the semi-direct sum of the solvable algebra
 with only $Q_\alpha ^i$ and $P_\mu$ which is in the first line of
(\ref{superPoincare}), the Lorentz-algebra $SO(3,1)$, and the
$R$-symmetry algebra $U(4)$.
\par
Now we look to extensions with
central charges. As explained in footnote~\ref{fn:central}, we
use this terminology to denote generators which appear in the
anticommutator of supersymmetries. These thus also contain generators which are neither the space-time
symmetry generators, nor singlets under space-time symmetries, \ie\
generators violating the Coleman--Mandula theorem \cite{ColemanMandula}.
We can
add generators to the anticommutator of the supersymmetries:
\tabcolsep 0pt
\begin{equation}\label{OSp116c}
\begin{array}{ccccc}
\left\{Q_\alpha ^i,Q_{\beta j}\right\}&=&\ft12\left((1-\gamma_5)\gamma^\mu\right) _{\alpha \beta}P_\mu \delta^i_j
&+&\ft12\left((1-\gamma_5)\gamma^\mu\right) _{\alpha \beta}Y_\mu{}^i{}_j \\
&& (4,1) &+&  (4,15)\\
\left\{Q_\alpha ^i,Q_\beta^j\right\}&=&\ft12(1-\gamma_5)_{\alpha \beta}R^{[ij]}&+&
 \ft12\left((1-\gamma_5)\gamma^{\mu \nu}\right)_{\alpha \beta} Z_{\mu \nu} ^{(ij)}\\
&& 2\ (1,6) &+&(6,10)\ ,
\end{array}
\end{equation}
\tabcolsep 6pt
where $[ij]$ denotes an antisymmetric tensor,
$(ij)$ a symmetric tensor and $Y$ is a traceless  tensor. $R$ is
complex while for $Z$ the complex conjugates are the selfdual and
antiselfdual components.
We have taken at the right hand side all possible terms consistent
with the symmetry, in total 136 bosonic generators.
We have given their content in representations of $SO(3,1)\times SU(4)$.
The algebra $OSp(1|16)$ has all the generators in  \eqn{OSp116c}.
In that superalgebra the 136 bosonic generators have non-trivial
commutation relations, namely those of $Sp(16)$, and the fermionic
generators
have the commutation relations with the bosonic ones
appropriate for a 16-representation. However, as
extension of the super-\Poin\ algebra, all these bosonic generators mutually
commute, and commute with the supersymmetries. Thus one may regard
that
algebra as a `contraction' of $OSp(1|16)$, where first
all the fermionic generators are scaled with $x$, the bosonic ones with
$x^2$,
and then the limit $x\rightarrow \infty$ is taken.
The full extended super-\Poin\
algebra is as before the semi-direct sum of this algebra with the $SO(3,1)$ Lorentz-generators
$M_{\mu \nu}$, and the $U(4)$ automorphism generators $U^i{}_j$. All generators have commutators with
$M_{\mu \nu}$ and $U^i{}_j$ as indicated by their indices.
\par
In a superconformal extension, we should also include the special
supersymmetries $S_\alpha ^i$, appearing in the commutator
$[K_\mu, Q_\a^i]=(\gamma_\mu)_\a^\b\,S_\b^i$. These have opposite
chirality as $Q^i$, thus $S^i=\ft12(1+\gamma_5)S^i$.
The usual superconformal algebra has
anticommutators between the fermionic generators:
\begin{eqnarray}
\left\{Q_\alpha ^i,Q_{\beta j}\right\}&=&\ft12\left((1-\gamma_5)\gamma^\mu\right) _{\alpha \beta}P_\mu
\delta^i_j
\nonumber\\
\left\{Q_\alpha ^i,S_{\beta j}\right\}&=&-\ft12(1-\gamma_5) _{\alpha
\beta}D\,
\delta^i_j -
\ft12\left((1-\gamma_5)\gamma^{\mu \nu}\right) _{\alpha \beta}M_{\mu \nu} \delta^i_j
+\left( 1-\gamma_5\right)_{\alpha \beta}U^i{}_j
\nonumber\\
\left\{S_\alpha ^i,S_{\beta j}\right\}&=&-\ft12\left((1+\gamma_5)\gamma^\mu\right) _{\alpha \beta}K_\mu
\delta^i_j\ .
\label{confd4N4}
\end{eqnarray}
Here $(P_\mu ,D,M_{\mu \nu},K_\mu )$ are the generators of the conformal
algebra $SO(4,2)$
and $U^i{}_j $ generate\footnote{In fact, this case is special
in that the trace of $U^i{}_j$ is not part of the simple algebra. It
can either be included as a central charge or as an automorphism.
In the first case, this means that the trace of $U$ is included in
(\ref{confd4N4}), but then it is central, \ie\ it commutes with the
supersymmetries, and we should thus write
\[
\left[ U^i{}_j,Q^k\right]= \delta^k_j Q^i- \ft14 \delta^i_j Q^k\ .
\]
Alternatively, the $U(1)$ factor does not commute
with the supersymmetries, \ie\ (\ref{superPoincare}) remains unchanged,
but then it should not appear in
(\ref{confd4N4}), where we should thus replace $U^i{}_j$ by
$U^i{}_j-\ft14\delta^i_j U^k{}_k$.}
 $SU(4)$. This superconformal algebra is $SU(2,2|4)$. It can be truncated to
the super-\Poin\ algebra given above by deleting the generators $S$, $D$ and $K$.
\par
Again we can consider a larger algebra with all generators allowed in
the supersymmetry commutators. It contains the anticommutator
\eqn{OSp116c} and a copy of it with $Q$ replaced with $S$ and all the
bosonic generators getting a partner in the same way as $K_\mu $ is
the partner of $P_\mu $ in \eqn{confd4N4}. The remaining anticommutators
are
\begin{eqnarray}
\left\{Q_\alpha ^i,S_{\beta j}\right\}&=&-\ft12(1-\gamma_5) _{\alpha \beta}D
\delta^i_j -
\ft12\left((1-\gamma_5)\gamma^{\mu \nu}\right) _{\alpha \beta}M_{\mu \nu} \delta^i_j
+\left( 1-\gamma_5\right)_{\alpha \beta}U^i{}_j
\nonumber\\
&&+\ft12\left((1-\gamma_5)\right) _{\alpha \beta}Y^{\prime\, i}{}_j
+\ft12\left((1-\gamma_5)\gamma^{\mu\nu}\right) _{\alpha \beta}
Y^{\prime\prime}_{\mu\nu}{}^i{}_j
\nonumber\\
\left\{Q_\alpha ^i,S_\beta^j\right\}&=&
\ft12\left((1-\gamma_5) \gamma^\mu\right) _{\alpha \beta}\left(R^\prime_\mu{}^{[ij]}
+ Z^\prime_\mu{} ^{(ij)}\right)\ .\label{QSD3ext}
\end{eqnarray}
The ones in the last anticommutator are complex.
The names have been chosen such that the generators with primes
combine with
those of $\{Q,Q\}$ without primes and their partners in $\{S,S\}$ in
representations
of $SO(4,2)$. In this way we have in the bosonic algebra
\begin{equation}
(15,1)(P,D,M,K) + [(1,15)+(1,1)](U)+ 2\,(6,6)(R)+(15,15)(Y)+(20,10)(Z)
\end{equation}
as decomposition under $SO(4,2)\times SU(4)$. The generators have
all non-trivial commutation relations, as they form the algebra $Sp(32)$.
The superalgebra is $OSp(1|32)$.
\par
Let us finish by giving an overview of all the algebras involved,
starting from $OSp(1|32)$.
\begin{equation}\label{fig:algebras}
\begin{array}{ccccc}
&\mbox{\bf Extended WV algebra}&&\mbox{\bf WV algebra}&\mbox{\bf
Supercharges}\\ &&&&\\ &\mbox{\rm M-algebra}&&&32\\ &&&&\\
&\uparrow \mbox{\rm\footnotesize contraction}&&&\\
\cline{1-1}&&&&\\ {a=0}&OSp(1|32)&\supset&SU(2,2|4)&32\\ &&&&\\
{\bf D3}&\cup&&\cup&\\ &&&&\\ {a=1}&\mbox{\rm Extended
Poincar\'e}&\supset&\mbox{\rm Poincar\'e}&16\\ &&&&\\ &\uparrow
\mbox{\rm\footnotesize contraction}&&\uparrow
\mbox{\rm\footnotesize contraction}&\\ \cline{1-1}&&&&\\
{a=0}&OSp(1|16)&\supset&OSp(4^*|4)&16\\ &&&&\\ {\bf D5\perp
D3}&\cup&&\cup&\\ &&&&\\ {a=1}&\mbox{\rm Extended
Poincar\'e}&\supset&\mbox{\rm Poincar\'e}&8\\ &&&&\\ &\uparrow
\mbox{\rm \footnotesize contraction}&&&\\ \cline{1-1}&&&&\\ &OSp(1|8)&&&8
\end{array}
\end{equation}
 The contraction of this algebra with a factor $x$
for the fermionic generators, a factor $x^2$ for the bosonic ones and
taking the limit $x\rightarrow \infty$ gives the $M$-algebra. Writing
its generators as representations of the 4 dimensional Lorentz group
and $U(4)$ we have the extended worldvolume algebra of the D3
background in the near-horizon case, containing the anticommutators
in (\ref{OSp116c}) and in (\ref{QSD3ext}). It has two important
subalgebras. One is the non-extended worldvolume superconformal
algebra $SU(2,2|4)$ containing the anticommutators in
(\ref{confd4N4}), the other is the extended super-\Poin\ algebra,
where the $S$-supersymmetries have been removed, together with parts
of other $SO(4,2)$ multiplets. The non-extended super-\Poin\ algebra
is a subalgebra of the superconformal non-extended algebra
as well as of the extended super-\Poin\ algebra. Both these \Poin\ algebras
are contractions of simple algebras, respectively $OSp(1|16)$ and
$OSp(4^*|4)$, which are near-horizon algebras for the theory we
consider in the paper, with a D5 worldvolume in a D3 background.
The constructions explained above can be repeated in this smaller
algebra, with a similar structure of subalgebras and contractions.
\par
Remark that by taking other decompositions of the 32-component
spinors we could, instead of going to $SU(2,2|4)$, also have gone to
{\it e.g.} $OSp(8|4)$ for the conformal
algebra in 3 dimensions, related to M2 or to $OSp(8^*|4)$ for 6 dimensions,
related to M5 \cite{m5tens}. The general setup of this article is
also applicable to these cases.


\end{document}